\documentclass[conference]{IEEEtran}
\IEEEoverridecommandlockouts
\usepackage{cite}
\usepackage{amsmath,amssymb,amsfonts}
\usepackage{algorithmic}
\usepackage{graphicx}
\usepackage{textcomp}
\usepackage{xcolor}
\usepackage{booktabs}
\usepackage{multirow}
\usepackage{multicol}
\usepackage{verbatim}
\usepackage{marvosym}

\usepackage[a4paper, total={184mm,239mm}]{geometry}   

\def\BibTeX{{\rm B\kern-.05em{\sc i\kern-.025em b}\kern-.08em
    T\kern-.1667em\lower.7ex\hbox{E}\kern-.125emX}}
\begin{document}

\title{MACO: Exploring GE\underline{M}M \underline{A}cceleration on a Loosely-\underline{Co}upled Multi-core Processor\\
}



\author{\IEEEauthorblockN{Bingcai Sui$^{1,2}$, Junzhong Shen$^{1,2}$, Caixia Sun$^{1,2}$, Junhui Wang$^{1,2}$, Zhong Zheng$^{1,2}$, Wei Guo$^{1,2}$\textsuperscript{\Letter}}
\IEEEauthorblockA{\textit{$^1$School of Computer, National University of Defense Technology} \\
\textit{$^2$Key Laboratory of Advanced Microprocessor Chips and Systems}\\
Changsha, China \\
\{bingcaisui, shenjunzhong, cxsun, wangjunhui, zheng\_zhong, wineer\_guowei\}@nudt.edu.cn}
}

\maketitle

\begin{abstract}


General-purpose processor vendors have integrated customized accelerator in their products due to the widespread use of General Matrix-Matrix Multiplication (GEMM) kernels. However, it remains a challenge to further improve the flexibility and scalability of these GEMM-enhanced processors to cater to the emerging large-scale GEMM workloads. In this paper we propose \textit{MACO}, a novel loosely-coupled multi-core general-purpose architecture optimized for GEMM-related applications. To enhance the programmability and flexibility of \textit{MACO}, the paper introduces a tile-based instruction set architecture. Additionally, the paper presents techniques such as hardware-assisted data prefetching and locking, and predictive address translation to further enhance the computational efficiency of \textit{MACO} for GEMM workloads. The experimental results demonstrate that \textit{MACO} exhibits good scalability, achieving an average computational efficiency of 90\% across multiple cores. Furthermore, evaluations on state-of-the-art deep neural networks show that \textit{MACO} can achieve up to 1.1 TFLOPS with 88\% computational efficiency, indicating its adaptivity to deep learning workloads.

\end{abstract}


\section{Introduction}

General Matrix-Matrix Multiplication (GEMM) is a critical building block for many domains including, but no limited to, high-performance computing (HPC), computer vision (CV), and natural language process (NLP). 
The demand for high-efficiency GEMM computations has driven the development of domain-specific architectures (DSA) such as Google's TPU~\cite{Jouppi2017} and Nvidia's tensor core, which are designed to meet the performance and energy-efficiency requirements.
To address the need for efficient AI application processing, CPU vendors like Intel, IBM, and ARM have started integrating deep learning-specific co-processors or execution units in their products. These solutions, referred to as GEMM-enhanced CPUs in the paper, extend the Instruction Set Architectures (ISAs) of Intel (AMX) and ARM (SME) to enable CPUs to execute GEMM workloads on customized execution units.

The paper classifies GEMM-enhanced CPUs into two categories: tightly-coupled architectures (TCA)~\cite{AMX, Jeong2021} and loosely-coupled architectures (LCA)~\cite{Genc2021, Lichtenau2022}. In TCA, the matrix accelerators are considered as an integral part of the CPU's pipeline. On the other hand, LCA designs treat the matrix accelerators as co-processors for the CPU core.
The authors highlight that while TCA offers advantages like reduced area budget and synchronization overhead between the CPU core and matrix accelerator, the performance of TCA solutions can be impacted due to resource contention between the CPU core and matrix accelerator.


LCA solutions provide several advantages compared to tightly-coupled architectures. One of the key benefits is the ease of use they offer. Loosely-coupled architectures simplify the design and implementation process, making it more straightforward to integrate and utilize components like the CPU and matrix accelerator.
Furthermore, these architectures excel in parallel computing, enabling efficient parallelization of tasks between the CPU and matrix accelerator. In this way, LCA solutions have a wider range of application scenarios, for example, when deploying recommended system on these architectures, we can offload top and bottom MLPs to the matrix engine leaving the CPU core free to run embedding lookups.

Overall, loosely-coupled architectures contribute to enhanced usability and facilitate parallel computing between the CPU and matrix accelerator, leading to improved performance in various applications.
However, the generality of the loosely-coupled architectures can be further improved, and this is the main focus of our paper. Gemmini~\cite{Genc2021}, a representative of loosely-coupled architectures, provides address translation support but does not consider the possible overhead of the accelerator in memory access caused by frequent cache misses when dealing with large-scale GEMM workloads. Additionally, Gemmini does not provide a specific solution for multi-process support and exception handling. Telum~\cite{Lichtenau2022} supports multi-processing and exception event handling, but the processes on multiple cores cannot run GEMM in parallel since only one process can occupy the shared AI accelerator at a time (via arbitration), causing frequent process switching and recovery. Moreover, although Gemmini and Telum both implement multi-core solutions, none of them provide details of mapping GEMM tasks on multiple cores, and the benefits of parallel computing on multiple cores are not being fully exploited.

To address the above issues, this work proposes \textit{MACO}, a loosely coupled multi-core general-purpose processor architecture with enhanced GEMM computation ability. 
Compared with Gemmini, \textit{MACO} 
provides better hardware support in multi-process execution, virtual-to-physical address translation, and exception event handling. Compared to Telum, all general-purpose cores of \textit{MACO} are accompanied by a separate matrix multiplication acceleration engine (\textit{MMAE}), allowing all cores to execute GEMM processes simultaneously, providing a higher degree of parallelism for GEMM. In addition, 
to improve \textit{MACO}'s usability and adaptability to various GEMM-relative applications, this work optimizes the architectural design from the following aspects:
(1) An instruction set extended from ARMv8 
is proposed to provide users with a variety of functions including GEMM computation, data migration, and data prefetch; (2) Efforts are made to enhance the architectural support for address translation, multi-process management and exception handling.
Moreover, We further explore the possibility of parallel computation of CPUs and MMAEs on \textit{MACO} for applications combined with both GEMM and non-GEMM workloads.
Finally, this work goes deeper than state-of-the-art Matrix-Multiplication-enhanced CPUs by presenting details of mapping scheme for GEMM workloads on multiple cores of \textit{MACO}. 
The main contributions of this work are summarized as follows:

\begin{itemize}
    \item We propose a novel loosely-coupled multi-core processor architecture named \textit{MACO}, which is featured by integrating multiple CPU+MMAEs (GEMM Acceleration Engine) as well as a highly scalable NOCs with cache coherence.

    \item An extended instruction set is proposed to improve the programmability and flexibility of \textit{MACO}, which also exposes \textit{MACO}'s rich functionality to users.

    \item We developed latency-hiding address translation technique based on page table address prediction to further improve GEMM performance on \textit{MACO}.

    
\end{itemize}

The experimental results indicate that \textit{MACO} has the capability to achieve a maximum throughput of 1.1 TFLOPS while maintaining a high computational efficiency of 88\%.

\section{Background}

\subsection{GEMM-enhanced CPUs}


\textbf{Tightly-coupled architectures}. Like Intel's Advanced Matrix Extensions (AMX)~\cite{AMX}, RASA~\cite{Jeong2021} which place MAE inside the CPU core, the main feature of tightly-coupled architecture (TCA) is regarding MAE as part of the CPU pipeline.The obvious advantage of TCA is that the MAE can share the resources of the CPU core, contributing to a reduction in the area overhead of the CPU chip.However, the execution procedures of the CPU and MAE are also tightly-coupled, meaning that both of them may suffer performance loss due to competition for resources (e.g MMU, LSU).

\textbf{Loosely-coupled architectures}. Representative examples for loosely-coupled architectures (LCA) include Telum and Gemmini, which consider MAE as a co-processor of the CPU core. In this scenario, since MAE has independent data paths and data load/store units (e.g DMA engine) for accessing last level cache or external memory, providing better opportunities for parallel computing between CPU and MAE. However, the main defects of LCA lie in the following aspects: (1) high synchronisation overhead between the CPU core and MAE; (2) MAE has difficulties in performing address translation independently; (3) process management and exception handling become challenges for the CPU core.

\subsection{Mapping Tile GEMM algorithm on Systolic Arrays}

\begin{figure}[!t]
    \centering
    \includegraphics[width=\linewidth]{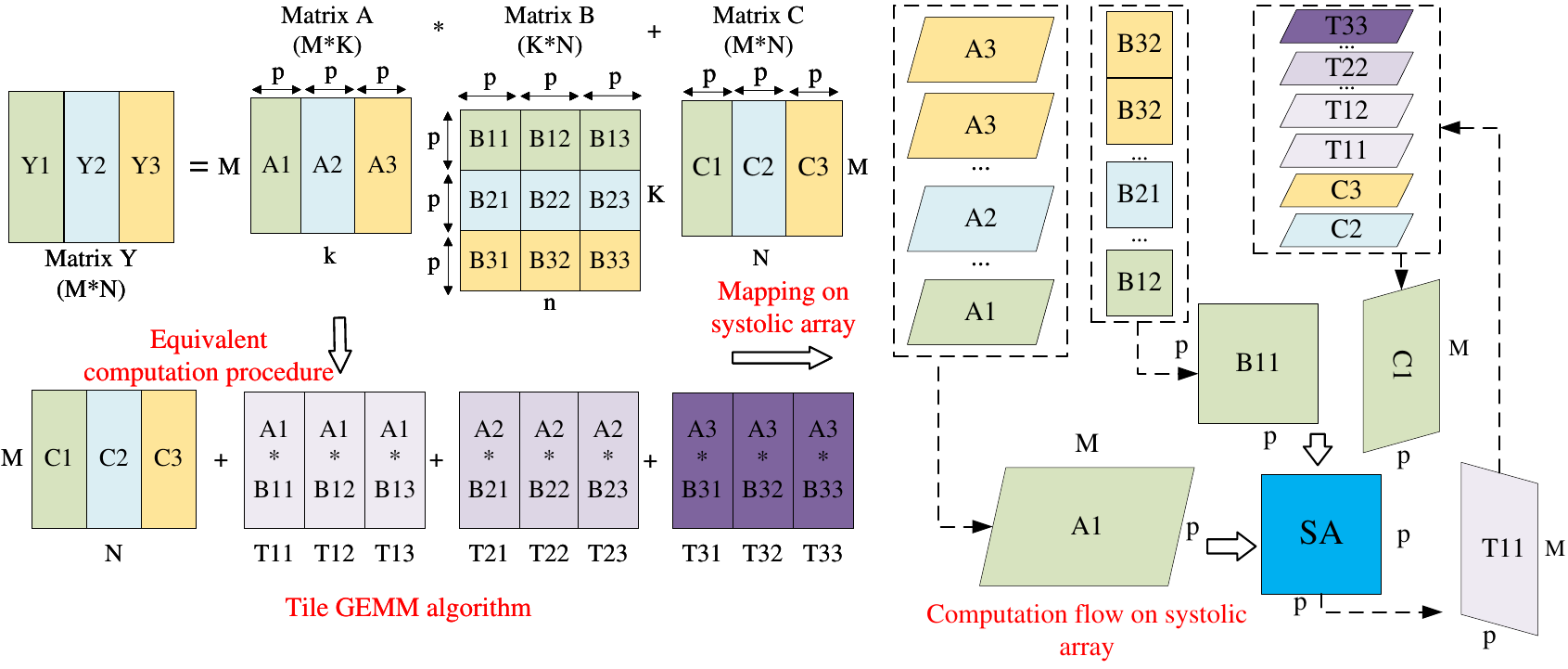}
    \vspace{-2.0em}
    \caption{Illustration of mapping tile GEMM algorithm on systolic array.}
    \vspace{-1.5em}
    \label{Fig:tile_gemm}
\end{figure}

Systolic arrays~\cite{kung1979systolic} have been successfully implemented in 
 various commercial products, including Intel's AMX~\cite{AMX}, Google's TPU~\cite{Jouppi2017}, and IBM's Telum~\cite{Lichtenau2022}. This is due to their advantages in simple construction, high concurrency, and efficient exploitation of the inherent data reuse of algorithms. 
Typical systolic arrays consist of many homogeneous processing elements (PE), each responsible for a MAC operation and interacting with each other through localized short concatenation lines for data interaction.

The left part of Fig.~\ref{Fig:tile_gemm} shows the classical tiled GEMM algorithm 
which breaks up the large matrices A, B and C into smaller sub-matrices for efficient computation. 
The right part of Fig.~\ref{Fig:tile_gemm} illustrates the schematic diagram of the equivalent computation procedure of tiled-GEMM algorithm mapped on a systolic array that employs the input-stationary data flow. 
In this case, the data of the sub-matrix B is pre-loaded and buffered within the systolic array before the data of sub-matrices A and C stream in. As each PE receives its data from the sub-matrices, it performs a local MAC operation and forwards the partial products either vertically (in the column direction) to its neighboring PE. 
During the computation, the partial products are temporarily stored in on-chip buffers, allowing for efficient data handling. These partial products are then loaded back into the systolic array to facilitate subsequent computations. This repeated procedure continues until the final results of the matrix multiplication operation are obtained.


\section{Architectural Design of \textit{MACO}}

\subsection{Overview}

\begin{figure*}[!t]
    \centering
    \includegraphics[height=5.5cm]{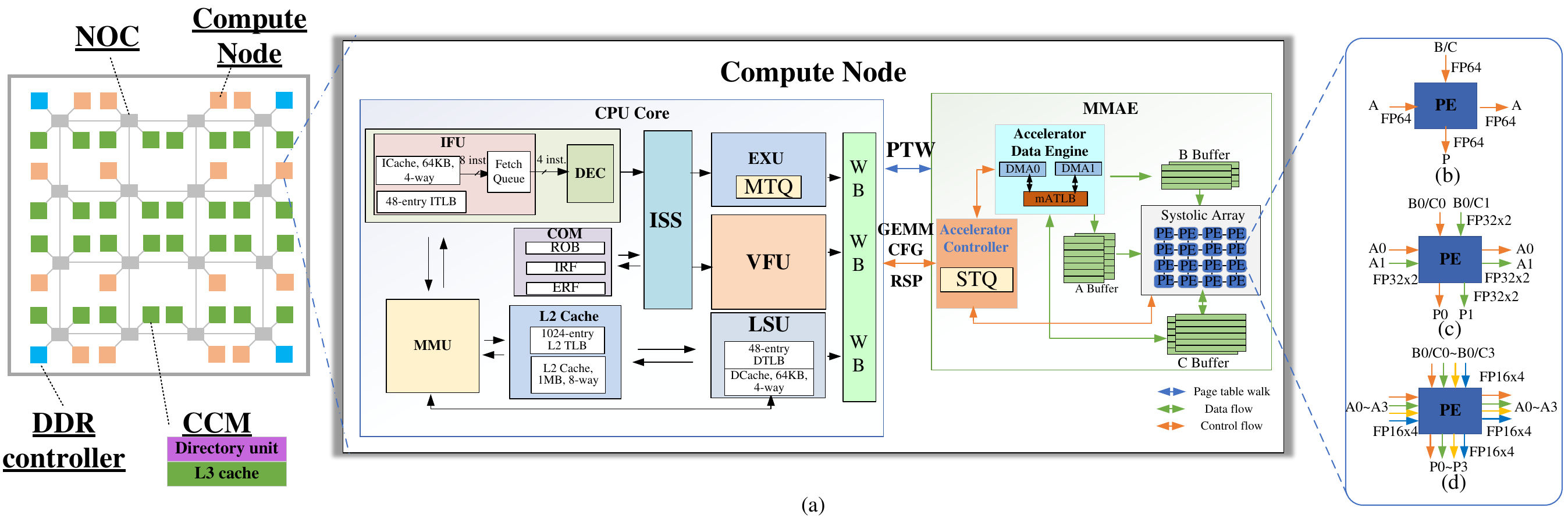}
     \vspace{-1.0em}
    \caption{Overview of \textit{MACO} architecture.}
    \vspace{-1.5em}
    \label{Fig:maco_arch}
\end{figure*} 


In Fig.~\ref{Fig:maco_arch}, \textit{MACO} consists of up to 16 homogeneous compute nodes interconnected by a network on chip (NOC). Each compute node is integrated with a general-purpose processor core that is associated with a matrix multiplication acceleration engine (\textit{MMAE}). The \textit{MMAE} shares the CPU core's shared TLB (sTLB) via customized interfaces. However, the CPU's L1 data cache and L2 cache are not accessible to \textit{MMAE}. Similarly, the CPU has no access to the associated \textit{MMAE}'s on-chip buffers. The CPU core is a 64-bit high-performance general-purpose processor core with a multi-issue superscalar architecture. Table~\ref{Tbl:MACO_AP} reports the architectural parameters of a CPU core. 

Form the right part of Fig.~\ref{Fig:maco_arch}(a), it can be seen that \textit{MMAE} is built on a 4×4 two-dimensional systolic array (SA) with integrated high-capacity buffers of 192KB for efficient data reuse. Its main function is to execute tile GEMM operations. More specifically, the integrated Accelerator Data Engine (ADE) is responsible for transferring data between L3 cache and on-chip buffers. And the Accelerator Controller (AC) functions by receiving configurations from the CPU and then scheduling SA , ADE and AC modules to complete the GEMM tasks.
Different from previous work, We further extend the classical dataflow of systolic array (see Fig.~\ref{Fig:tile_gemm}) to support SIMD-like compute modes including 2-way FP32 (Fig.~\ref{Fig:maco_arch}(c)) and 4-way FP16 (Fig.~\ref{Fig:maco_arch}(d)) parallel computations. 
In addition, by integrating powerful DMA engines, \textit{MMAE} can carry out high-capacity data initialization and data migration without disturbing the CPU core. More importantly, the above procedures are fully programmable via the proposed instructions (details follow), which effectively improves the  flexibility of \textit{MMAE}.

The NOC prototype is a classical 2D mesh network of size 4×4, and each node of NOC provides multiple interfaces for a compute node, cache coherence manager (CCM), external memory controller (optional), or I/O controller (optional). NOC can provide up to 128 GB/s memory bandwidth for each compute node (bidirectional read/write bandwidth, 256-bit@2GHz). NOC supports X-Y routing algorithm and virtual channels flow control, providing reliable data transfer between source and destination nodes. The L3 cache (also named system cache) is distributed among all CCMs and shared by all compute nodes. CCM implements a directory-based cache consistency protocol, which functions by tracking and recording the data states (based on MOESI protocol) inside the L3 cache and maintaining data consistency between compute nodes across the chip.

\begin{table}[!tbp]
\caption{Architectural Parameters of a CPU Core}
\vspace{-0.7em}
\label{Tbl:MACO_AP}
\centering
\begin{tabular}{l| c} \toprule
Architectural Parameters     & Value                        \\ \midrule
instruction width            & 64-bit                           \\ 
data bus width               & 256-bit, CHI protocol         \\
instruction fetch width      & 128-bit                           \\
pipeline stages              & 12+                          \\
instruction execution order  & out-of-order                 \\
multi-issue ability          & four-issue                   \\
L1 Instruction Cache(ICache) & 48KB, four-way set associate \\
L1 Date Cache (Dcache)       & 48KB, four-way set associate \\
L2 Cache                     & 512 KB, private   \\ 
L1 ITLB/DTLB                 & 48 entries, fully associate   \\
L2 TLB                       & 1024 entries, fully associate  \\ \bottomrule
\end{tabular}
\vspace{-1.5em}
\end{table}

\subsection{Matrix Processing Assist Instruction Set}


We propose a new non-privileged instruction set called Matrix Processing Assist (\textit{MPAIS}), which extends the ARMv8 instruction set architecture (ISA). \textit{MPAIS} includes three key GEMM-related functions: data migration, tile GEMM computation, and task management. Table~\ref{MPA} details the instructions of \textit{MPAIS} and their specified functions. To enable users to commit their tile GEMM tasks to the \textit{MMAE}, we have implemented GEMM computing instructions. Users must allocate six successive general registers (i.e., Rn, Rn+1, ..., Rn+5) for storing GEMM-related parameters before using the MA\_CFG instruction. The MA\_CFG instruction consists of a series of micro-operations (mops), such as requesting an available entry of \textit{Master Task Queue} (Details in Section III.C)  and sending the buffered parameters to the \textit{MMAE}. If an \textit{MTQ} entry is successfully allocated, the identifier of the entry (named MAID) will be stored in the destination register \textit{Rd}.

We have also designed data migration instructions (MA\_MOVE, MA\_INIT, and MA\_STASH) to utilize the DMA engines of the \textit{MMAE} for fast data transfer or initialization. All instructions have the same execution flow as MA\_CFG, but the parameters stored in the allocated registers Rn$-$Rn+5 differ. The \textit{MMAE} can decode the parameters and executes corresponding operations independently. 

Task management instructions MA\_READ and MA\_STATE can help users to obtain the execution states of their GEMM tasks via the previously stored MAID (i.e the destination register of MA\_CFG). 
Both MA\_STATE and MA\_READ instructions can be used to query the execution states of an entry of \textit{MTQ}. MA\_STATE differs from MA\_READ by an additional "release" operation on the queried entry.
The obtained information is stored in the \textit{Rd} register specified by the instructions. Additionally, users can use MA\_CLEAR to clear the entry of \textit{MTQ} (specified by the MAID stored in Rd register) if exception events occur during task execution.


\begin{table}[!tbp]
\caption{Illustrations of the proposed instruction set.}
\vspace{-2em}
\begin{center}
\label{MPA}
\resizebox{1\columnwidth}{!}{
\begin{tabular}{c|c|c|c}
\toprule
Functions & Instructions & Description & Usage \\
\hline
 & \multirow{2}{*}{MA\_MOVE}  & Copy data from source &  \multirow{2}{*}{MA\_MOVE Rd, Rn} \\ 
 & & address to destination address. &  \\ 
\cline{2-4} 
Data  & \multirow{2}{*}{MA\_INIT} & Set data in destination & \multirow{2}{*}{MA\_INIT Rd, Rn} \\ 
migration & & space to zeros. & \\ 
\cline{2-4} 
 & \multirow{2}{*}{MA\_STASH} & Perform data prefetch from the & \multirow{2}{*}{MA\_STASH Rd, Rn} \\
 & & external memory to L3 cache. & \\
\hline
GEMM  & \multirow{2}{*}{MA\_CFG} & Request an \textit{MTQ} entry for & \multirow{2}{*}{MA\_CFG Rd, Rn} \\
computing &  & executing a GEMM task. & \\
\hline
 & \multirow{2}{*}{MA\_READ} & Obtain the execution state of  & \multirow{2}{*}{MA\_READ Rd, Rn} \\ 
 &  & a certain GEMM task. & \\ 
\cline{2-4}
Task  & & Obtain execution state of a certain & \\ 
management & MA\_STATE & GEMM task and release & MA\_STATE Rd, Rn \\ 
 &  &  the occupied MTQ entry. & \\ 
\cline{2-4}
 & MA\_CLEAR & Clear a certain \textit{MTQ} entry. & MA\_CLEAR, Rn \\        
\bottomrule  
\end{tabular}
}
\vspace{-1.5em}
\label{tab1}
\end{center}
\end{table}

\subsection{Multi-process Management}

\begin{table}[!t]
\centering
\caption{Details of an \textit{MTQ} entry.}
\label{Tbl:MTQ_entry}
\begin{tabular}{l|l} \toprule
Field         & Description                                                            \\ \midrule
Valid         & Indicate whether the entry is allocated.                              \\ 
Done          & Indicate whether the task is completed.                                \\
ASID          & Process identifier.                                                    \\
exception\_en & Indicate exception occurs during \textit{MMAE}'s task execution. \\ 
exception\_type & Specific type of an exception event. \\ \bottomrule                                 
\end{tabular}
\vspace{-1.0em}
\end{table}


\begin{figure}[!t]
    \centering \includegraphics[width=\linewidth]{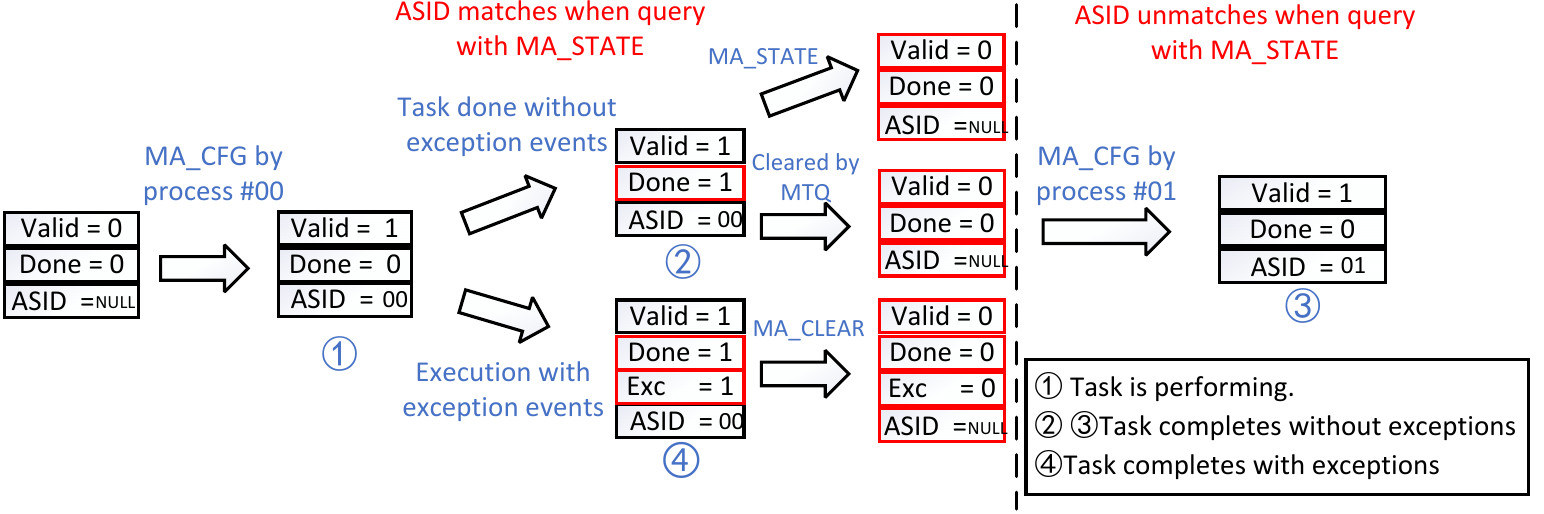}
    \vspace{-1.5em}
    \caption{State transition diagram of an \textit{MTQ} entry.}
    \vspace{-2.0em}
\label{Fig:MTQ_state}
\end{figure}

We have integrated a \textit{Master Task Queue} (\textit{MTQ}) and \textit{Slave Task Queue} (\textit{STQ}) in each CPU and \textit{MMAE} to timely record the state of all GEMM process, respectively. Each \textit{MTQ} has multiple entries, each of which can record the execution state of a GEMM process independently. Table~\ref{Tbl:MTQ_entry} shows the details of an \textit{MTQ} entry. It can be seen that an \textit{MTQ} entry can provide rich information including the process identifier, task execution state and exception events. 
 The functions of \textit{STQ} include: receiving parameters of a GEMM task from the CPU core (identified by the same MAID), parsing parameters and saving them at its local registers, monitoring execution states of other components of \textit{MMAE} (e.g. DMA engines, systolic arrays), and responding the status of the GEMM task to the corresponding \textit{MTQ} entry. 

For GEMM workloads, the CPU core would send the GEMM relative parameters of the task to the \textit{MMAE}, thereby the \textit{STQ} entry specified by the MAID would receive and then buffer the configuration information locally. The buffered tasks in the \textit{STQ} entries will be automatically executed when the active 
\textit{STQ} entry has completed its task.
Fig.~\ref{Fig:MTQ_state} shows the state transition diagram of an \textit{MTQ} entry when process switch occur.
It can be seen that we can combine the values of "Done" and "ASID" from the specified \textit{MTQ} entry to determine the execution state of the process, even thought the entry has been occupied by other process (see state 3).
For state 4, since a GEMM task may be automatically terminated by the \textit{MMAE} if there are exception events during task execution, users have to do further check to determine the accurate type of the exception events. Note that both \textit{MTQ} and \textit{STQ} will not affected by process switching, thereby we can obtain reliable information associated with all the processes from their corresponding allocated \textit{MTQ} entries.

\section{Implementation Details of \textit{MACO}}

\subsection{Predictive Address Translation}

\begin{figure}[!t]
    \centering
    \includegraphics[width=0.95\linewidth]{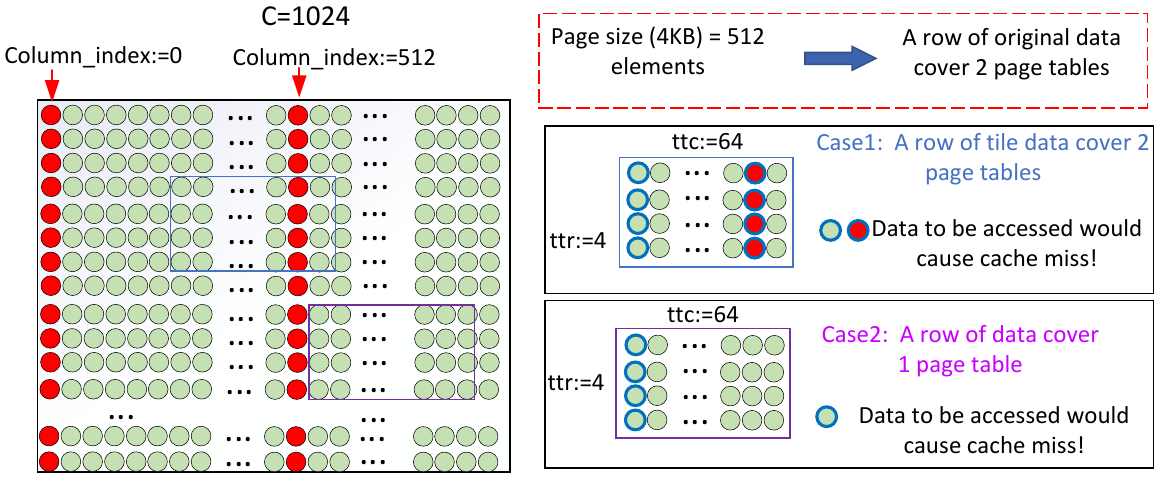}
    \vspace{-0.8em}
    \caption{Basics of page table address prediction.}
    \vspace{-1.0em}
    \label{Fig:mtlb}
\end{figure}

Fig.~\ref{Fig:mtlb} presents a simple example to illustrate 
how to determine the virtual addresses that would cause cache misses.
Note that parameters such as the number of columns of the original matrix ($C$), the tile size (i.e. $<Tr, Tc>$), and the page table size are configured to the \textit{MMAE} in advance. In Fig.~\ref{Fig:mtlb}, we assume the page table size is 4KB, and the original matrix consists of 1024 elements in FP64 precision (8KB in total), meaning that each row of the original matrix data are mapped to two page tables. As it can be seen that the elements identified by red circles represent the first data located at each page table, and once the location of the tile data in the original matrix is known, we can determine whether the data to be accessed would cause cache miss. Based on this observation, 
we design a module named \textit{mATLB} to generate multiple virtual addresses in advance, then sends them to the CPU core's memory manage unit (MMU) to perform page table walk. After a period of time, the returned \textit{ATLB} entries (including translated physical address) would be stored in the local buffers of \textit{mATLB}. Each entry would be accessed by the DMA engines then provide physical address for their memory access requests, and each entry would be removed from the buffer once it fails to match the current virtual address. In this way, the overhead of cache misses  would be perfectly hidden by pre-performed page table walk. 


\subsection{Mapping Real-world GEMM$^+$ Workloads on \textit{MACO}}

\begin{figure}[!t]
    \centering
    \includegraphics[width=0.88\linewidth]{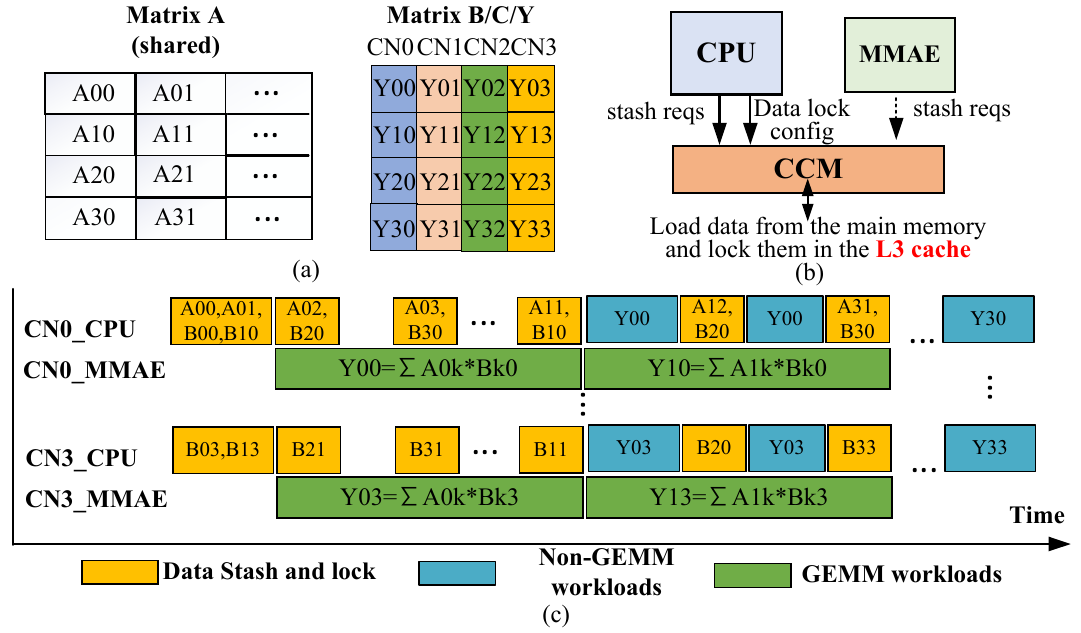}
    \vspace{-0.95em}
    \caption{Mapping schemes of GEMM$^+$ workloads on \textit{MACO}.} 
    \vspace{-1.5em}
    \label{Fig:maco_mapping}
\end{figure}

In real-world applications involving CNNs, RNNs and transformer-based models, it is common to follow GEMM-based workloads (such as convolutional layers, fully-connected layers, and attention layers) with non-GEMM but complex workloads, including normalization, activation, and softmax functions. 
This work present a novel and efficient scheme for mapping these combined GEMM (GEMM$^+$ for short) workloads onto \textit{MACO}. 
Fig.~\ref{Fig:maco_mapping}(b) provides an illustration of the data prefetch and locking procedure. Both the CPU and MMAE can issue "stash" requests to the CCM, enabling the prefetching of sub-matrices into the L3 Cache. Subsequently, the CPU can generate configurations lock the data in the L3 cache via the CCM. The advantages of data prefetching and locking two-fold: first, the memory access efficiency of \textit{MMAE} could be guaranteed since there are no page misses occur during Page Table Walk (PTW); secondly, if the result tiles of the GEMM workloads also reside in the L3 cache, the CPU can perform subsequent non-GEMM workloads without incurring data misses. 
Fig.~\ref{Fig:maco_mapping}(c) illustrates the timing graph depicting the mapping of GEMM$^+$ workloads onto the four compute nodes of \textit{MACO}. As shown that we achieve achieve parallelization by tiling the original matrices and efficiently allocating the resulting sub-matrices to the compute nodes (CNs). (see Fig.~\ref{Fig:maco_mapping}(a)). 

\section{Evaluations}

\subsection{Experimental Settings}




To evaluate the \textit{MACO} architecture, we performed both ASIC and FPGA flows.
In the ASIC flow, we ensured that the timing requirements of the CPU cores, MMAEs, and NOC were met at frequencies of 2.2 GHz, 2.5 GHz, and 2.0 GHz, respectively, after Placement and Routing. These timings were achieved using a 12 nm library.
In the FPGA flow, the RTL code of MACO underwent FPGA synthesis and layout using Vivado Design Suite 2022.2. The experiments were conducted on the Xilinx VCU440 FPGA, and the implemented \textit{MACO} design was clocked at 50 MHz.
On the FPGA platform, we successfully ran a modified Linux operating system on \textit{MACO}. 
Note that GEMM workloads of various sizes used for evaluation were obtained from an open-source software package~\cite{hpl}.

\subsection{Experimental Results}



\subsubsection{Evaluations on Area and Power}

\begin{table}[]
\caption{Comparisons of the CPU core and \textit{MMAE}. }
\vspace{-1em}
\label{area_power}
\begin{tabular}{l|c|c|c|c|c} \hline
     & Freq   & Area & Power & \multirow{2}{*}{FMACs} & Peak Perf$^{\mathrm{a}}$  \\ 
     & (GHz)  & (mm$^2)$ &  (W) &   & (GFLOPS)  \\ \hline 
CPU              & 2.2  & 6.25 & 2.0     & 8          & 35.2(FP64)/71(FP32) \\ \hline 
\multirow{2}{*}{\textit{MMAE}}        & \multirow{2}{*}{2.5}  & \multirow{2}{*}{\ \ \,1.58$^{\mathrm{b}}$ } & \multirow{2}{*}{1.5}   & \multirow{2}{*}{16}         & 80(FP64)/160(FP32)/ \\  
                     &      &                     &       &  &320(FP16) \\ \hline
\multicolumn{6}{l}{$^{\mathrm{a}}$ Theoretical peak performance, calculated by $2\times Freq \times FMACs $.} \\
\multicolumn{6}{l}{$^{\mathrm{b}}$ Area breakdown: Buffers: 36.7\%, SA: 24.7\%, AC: 23.4\%, ADE: 15.8\%.}
\end{tabular}
\vspace{-2.0em}
\end{table}

Table~\ref{area_power} compares the area and power consumption of a single CPU core and \textit{MMAE}.
It can be seen that the area of \textit{MMAE} is only  25\% of the size of CPU core, but the peak performance in GFLOPS of \textit{MMAE} is over 2$\times$ of that of CPU core. As a result, \textit{MMAE} can obtain a much higher (9$\times$) area efficiency (GFLOPS/mm$2$) than CPU. In addition, The power consumption of \textit{MMAE} is 25\% lower than CPU, contributing to a 2$\times$ theoretical computation efficiency (GFLOPS/W) than CPU. It can be concluded that we have effectively extended the GEMM computation power of the CPU at a smaller cost of area and power consumption.


\subsubsection{Evaluations on Address Translation}

\begin{figure}[!t]
    \centering
    \includegraphics[width=0.8\linewidth]{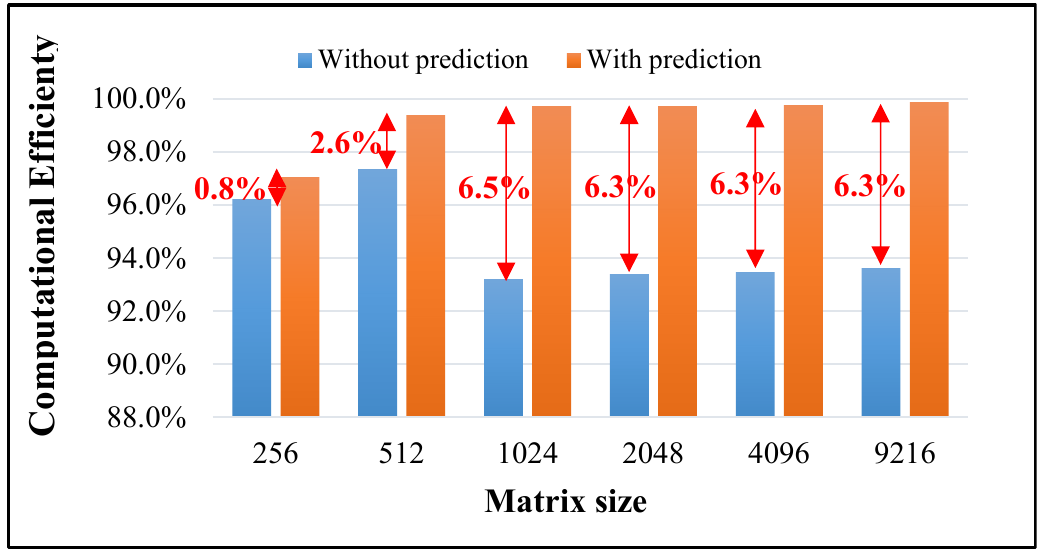}
    \vspace{-0.5em}
    \caption{Performance of MACO with/without page table prediction.}
    \vspace{-1.0em}
    \label{Fig:exp_mtlb}
\end{figure}


To evaluate the effectiveness of the proposed address translation technique, we conducted experiments to measure the sustainable performance of \textit{MACO} with and without prediction on address translation. In these experiments, we maintained a uniform page size and tiling size for all cases:  4KB pages and first-level tiling of $<Tr, Tc> = <1024, 1024>$, along with second-level tiling of $<ttr, ttc> = <64, 64>$. For simplicity, only one compute node was involved in these tests.

As depicted in Fig.~\ref{Fig:exp_mtlb}, it is evident that \textit{MACO} achieves higher computational efficiency when performing address translation in advance. Note that computational efficiency is calculated by the ratio of measured performance to theoretical peak performance in GFLOPS. The performance gap reaches a maximum of 6.5\% with a matrix size of 1024. However, for matrix sizes smaller than 512, the performance gains are not significant (which is expected), amounting to less than 2\%. This can be attributed to the fact that in those cases, the matrices fail to cover multiple page tables, resulting in fewer cache misses during the memory access of the MMAE. Consequently, the predictive address translation yields limited performance improvement in such scenarios.

\subsubsection{Scalability}

\begin{figure}[!t]
    \centering
    \includegraphics[width=\linewidth]{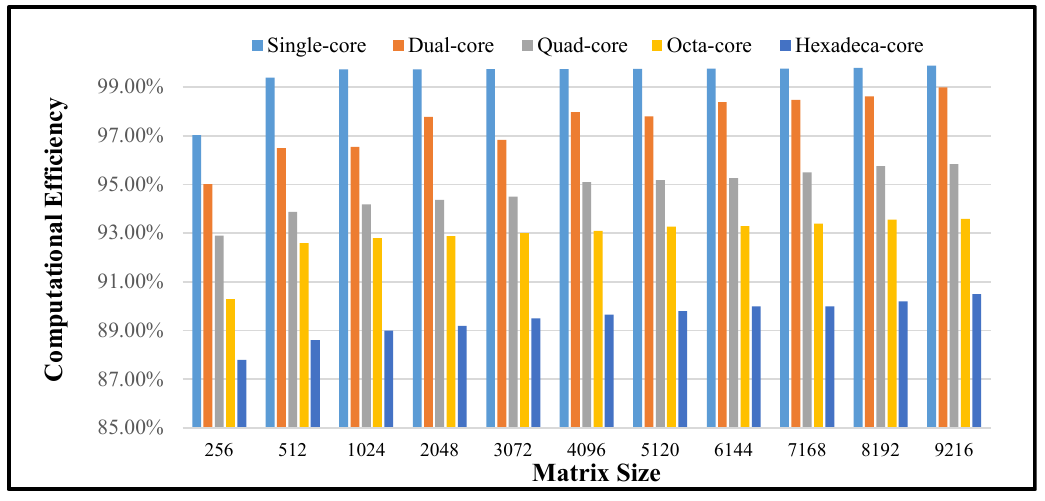}
    \vspace{-1.9em}
    \caption{Scalability of \textit{MACO}.}
    \vspace{-1.9em}
    \label{Fig:exp_scalability}
\end{figure}


To assess the scalability of \textit{MACO}, we conducted throughput tests using various GEMM workloads and varying the number of compute nodes (2, 4, 8, 16) involved. Each compute node was assigned an independent GEMM workload, with no inter-node interaction.
The results, as depicted in Fig.~\ref{Fig:exp_scalability}, indicate the average computational efficiency per compute node (y-axis) across different matrix sizes (x-axis). It is observed that with an increase in the number of compute nodes, \textit{MACO} experiences an average performance loss of 10\%. This can be attributed to the NOC being unable to meet the bandwidth requirements of all compute nodes working in parallel.
Despite the performance loss, \textit{MACO} still achieves an approximate computation efficiency of 90\% for all test cases. This result demonstrates that the proposed data prefetch and predictive address translation techniques effectively reduce the memory access overhead. Hence, we can conclude that \textit{MACO} exhibits good scalability in parallel computing with multiple compute nodes.

\subsubsection{Comparisons with state-of-the-art}

\begin{figure}[!t]
    \centering
    \includegraphics[width=0.8\linewidth]{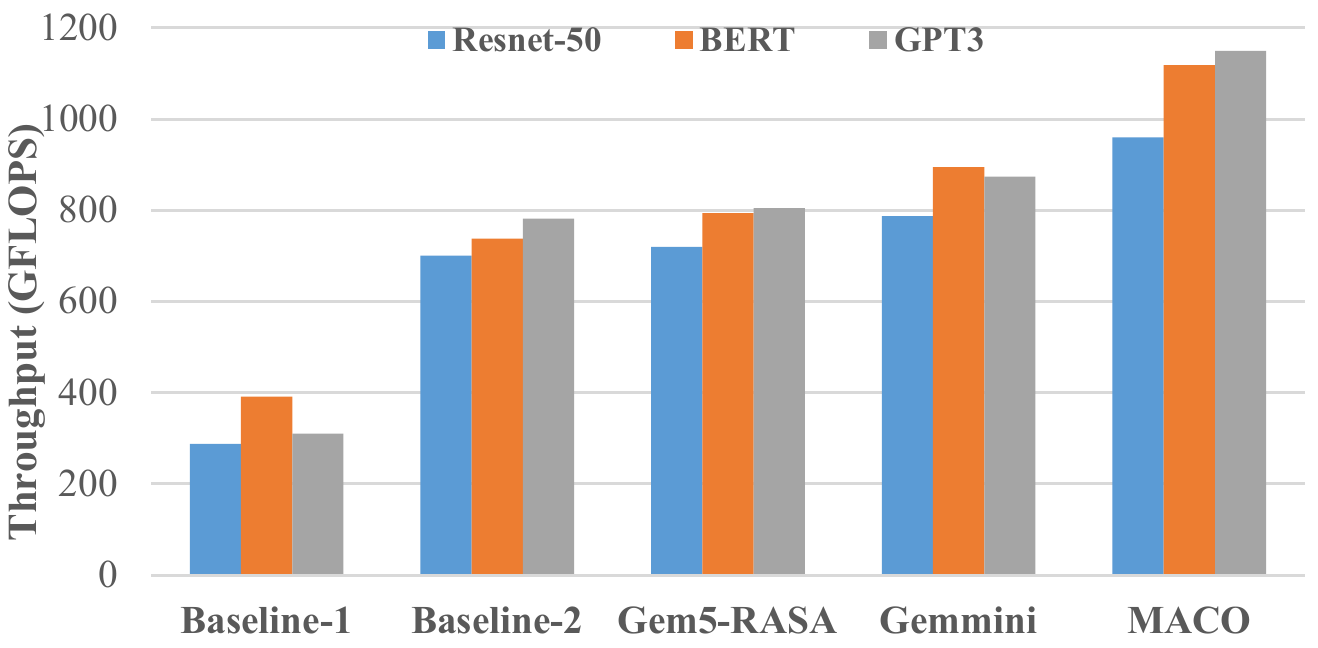}
    \vspace{-1.0em}
    \caption{Comparisons with state-of-the-art solutions with DL workloads.}
    \vspace{-1.5em}
    \label{Fig:maco_dl}
\end{figure}

To validate the superiority of \textit{MACO}, we conducted experiments using real-world deep learning (DL) workloads. The benchmarks selected for this study were Resnet-50~\cite{he2016deep}, BERT~\cite{devlin2018bert}, and GPT3~\cite{brown2020language}, all of which are used for inference purposes and employ FP32 precision. We compared \textit{MACO} against four counterpart solutions, including (1) Baseline-1, \textit{MACO} with CPU-only; (2) Baseline-2 \textit{MACO} with MMAE, but without applying mapping scheme illustrated in Section IV.B; (3) A modified version of MacSim with similar configurations to RASA~\cite{Jeong2021}; (4) Gemmini. To ensure a fair comparison, we configured all solutions with the same number of processing elements (16$\times$16). As illustrated in Fig.~\ref{Fig:maco_dl}, \textit{MACO} outperformed the other solutions across all benchmarks. More specifically, \textit{MACO} achieves an average performance gain  of 1.35x and 1.30x over RASA and Gemmini, respectively. When compared to Baseline-1 and Baseline-3, \textit{MACO} demonstrated remarkable performance improvements of 3.30x and 1.45x, respectively. In addition, \textit{MACO} can achieve up to 1.1 TFLOPS with 88\% computational efficiency, indicating its adaptivity to deep learning workloads.

\section{Related Work}

In the area of CPUs with efficient support for running dense or sparse GEMM operations, there are several related works that have been proposed.

For dense GEMM support on CPUs, ZCOMP~\cite{akin2019zcomp} introduces a vector Instruction Set Architecture (ISA) extension that reduces cross-layer communication in Deep Neural Networks (DNNs). This extension aims to improve performance by optimizing the data movement between different layers of the DNN.
RASA~\cite{Jeong2021} proposes control and data optimizations for CPU matrix engines to improve utilization through efficient pipelining and overlap. It divides matrix multiplication into different sub-stages on the systolic array and introduces optimizations with pipelining and overlapping to maximize throughput.

In terms of sparse GEMM support on CPUs, much of the related work focuses on accelerating Deep Neural Networks (DNNs). SAVE~\cite{gong2020save} is a sparsity-aware CPU vector engine that skips redundant computations in sparse GEMM operations, thereby accelerating sparse DNN and high-performance computing workloads.
SparCE~\cite{sen2018sparce} aims to increase the utilization of vector engines by tracking general-purpose registers with zeros. It skips ineffective code blocks based on sparse input by annotating skippable code blocks in software and testing conditions in hardware. This approach requires hardware-software co-design and primarily focuses on scalar code.
VEGETA~\cite{jeong2023vegeta} presents a set of ISA and microarchitecture extensions over dense matrix engines to support flexible structured sparsity for CPUs. It enables programmable support for diverse deep learning models with varying degrees of sparsity.


\section{Conclusion}


In this paper, we propose \textit{MACO}, an innovative loosely-coupled multi-core general-purpose architecture specifically designed for GEMM and related applications. In addition, to enhance \textit{MACO}'s programmability and flexibility, we further propose a tile-based instruction set \textit{MPAIS}. 
Moreover, an efficient mapping scheme for GEMM$^+$ workloads on \textit{MACO} is proposed to exhibit the parallel computing power of \textit{MACO}.
The aforementioned contributions collectively serve as a promising foundation for future architectural advancements in the realm of GEMM-enhanced general-purpose processors.

\section*{Acknowledgment}

 This work was supported by and National Natural Science Foundation of China (Grant No.U23A20301, No.62202481), TDRCJH program (Grant No.22-TDRCJH-02-006), KJWPDL program (Grant No.2022-KJWPDL-009), and NUDT Foundation (No.ZK2023-16). Bingcai Sui and Junzhong Shen are co-first authors of this paper. Wei Guo is the corresponding author of this paper.


\bibliographystyle{IEEEtran}
\bibliography{IEEEabrv,mybib1}

\end{document}